\documentstyle{article}
\newenvironment{quotes}{%
\begin{quote}
\small}
{\end{quote}
\normalsize}
\begin{document}
\begin{titlepage}

\begin{center}
\footnotesize
Forthcoming in {\it Mind}, July 1994.
\vspace{12pt}

\Large
{\huge {\it A Neglected Route to Realism about Quantum Mechanics}}\footnote{
An
early version of this paper was read at the AAHPSSS Conference
in Melbourne in May, 1988. Many people have helped me with comments
and discussion since then, and I am especially indebted to Jeremy
Butterfield, Peter Menzies and Jack Smart. I am also grateful
for research funding from the Australian Research Council.%
}

\vspace{12pt}\LARGE
{\sc Huw Price}

\vspace{9pt}\small
School of Philosophy\\University of Sydney\\Australia 2006

\vspace{3pt}
[email: huw@extro.su.edu.au]

\vspace{12pt}
\vspace{24pt}
{\bf {\large ABSTRACT}}

\end{center}%

\small\setlength{\parindent}{0pt}
Bell's Theorem depends on the assumption that hidden variables
 are not influenced by future measurement settings, and it is
widely recognised that some of the puzzling features of quantum
mechanics could be explained if this assumption were invalid.
The suggestion has generally been regarded as outlandish,
however, even by the taxed standards of the discipline.
(Bell himself thought that it led to fatalism.) This paper
 argues that there is surprisingly little justification
for this reaction. Using an informal model of the Bell correlations,
the first part of the paper shows that the arguments against
advanced action in QM, where not simply invalid, are easily
evaded on good physical grounds. In the absence of such objections
the approach has striking theoretical advantages, especially in
avoiding the apparent conflict between Bell's Theorem and special relativity.

\setlength{\parindent}{12pt}%
The second part of the paper considers the broader question
as to why advanced action seems so counterintuitive, by
investigating the origins of our ordinary intuitions about
 causal asymmetry. It is argued that the view that the past
does not depend on the future is largely anthropocentric, a
kind of projection of our own temporal asymmetry. Many physicists
have also reached this conclusion, but have thought that if
causation has no objective direction, there is no objective content
to an advanced action interpretation of QM. This turns out to
be a mistake. From the ordinary subjective perspective, we
can distinguish two sorts of objective world: one ``looks as if''
 it contains only forward causation, whereas the other ``looks as if''
 it involves a mix of backward and forward causation.
 This clarifies the objective core of an advanced action
interpretation of QM, and shows that there is an independent
 symmetry argument in favour of the approach.

\end{titlepage}

\setlength{\parindent}{0pt}
The most profound conceptual difficulties
of quantum mechanics are those that stem from the work of J.\ S.\ Bell in the
mid-1960s.\footnote{% Originally
Bell (1964). Bell's papers on the subject are collected in his
(1987).}%
As Bell's Theorem became well known, its author was often asked
to survey the state of the subject, particularly in the light
of his own contribution. He would typically conclude such a lecture
by listing what he saw as possible responses to the difficulties,
indicating in each case what he took to be the physical or philosophical
objections to the response concerned. His intuitions in this
respect were somewhat unfashionably realist---like Einstein,
he disliked the common view that quantum mechanics requires us
to abandon the classical idea of a world existing independently
of our observations. He therefore appreciated the irony in the
fact that from this realist standpoint, his own work seemed to
indicate that there are objective non-local connections in the
world, in violation of the spirit of Einstein's theory of special
relativity. As he puts it in one such discussion,

\setlength{\parindent}{12pt}%
\setlength{\parskip}{0pt}

\begin{quotes}%
the
cheapest resolution is something like going back to relativity
as it was before Einstein, when people like Lorentz and Poincar\'{e}
thought that there was an aether---a preferred frame of reference---but
that our measuring instruments were distorted by motion in such
a way that we could not detect motion through the aether. Now,
in that way you can imagine that there is a preferred frame of
reference, and in this preferred frame of reference things do
go faster than light.\footnote{In
Davies and Brown (1986), pp.\ 48-9. The irony actually runs deeper
than this, for Bell's Theorem seems to undercut Einstein's strongest
argument in favour of his view that there is more to reality
than quantum mechanics describes; more on this in section 1 below.
}%

\end{quotes}%
\setlength{\parindent}{12pt}%
\setlength{\parskip}{0pt}

 Bell reached this conclusion with considerable regret, of course,
and would often note that there is one way to save both locality
and realism. Bell's Theorem requires the assumption that the
properties of a system at a time are statistically independent
of the nature of any measurements that may be made on that system
in the future---``hidden variables are independent of later measurement
settings'', to put it in the jargon. Bell saw that in principle
one might defend a local (and hence special relativity friendly)
Einsteinian realism by giving up this %
{\it Independence Assumption}. He found this solution even less
attractive than that of challenging special relativity, however,
for he took it to entail that there could be no free will. As
he puts it, in the analysis leading to Bell's Theorem

\begin{quotes}

it is assumed that free will is genuine, and as a result of that
one finds that the intervention of the experimenter at one point
has to have consequences at a remote point, in a way that influences
restricted by the finite velocity of light would not permit.
If the experimenter is not free to make this intervention, if
that also is determined in advance, the difficulty disappears.
(Davies and Brown, 1986, p.\ 47)
\end{quotes}

It is surprising that philosophers have not paid
more attention to these remarks. In effect, Bell is telling us
that Nature has offered us a metaphysical choice of an almost
Faustian character. We may choose to enjoy the metaphysical good
life in quantum mechanics, keeping locality, realism, and special
relativity---but only so long as we are prepared to surrender
our belief in free will! The philosophical interest of the case
is hardly diminished by the fact that Bell himself preferred
to decline the temptation. It would be fascinating enough if
our philosophical engagement were merely that of spectators,
an audience to Bell's Faust. As it is, of course, the same metaphysical
offer is extended to all of us, and many philosophers may feel
that Bell was wrong to refuse. Some will have long since concluded
that there is no such thing as free will, and might thus take
the view that Nature is offering a very attractive free lunch.
Others have become adept at juggling free will (or some acceptable
substitute) and determinism, and hence might hope to take advantage
of the offer at very little real cost. (With respect, after all,
who is Bell to tell us what is incompatible with free will?)
Even if Bell is right, and it does come down to a choice between
a relativistically acceptable realism and free will, we might
feel that Bell simply makes the wrong choice---what we should
say, as honest empiricists, is simply that science has revealed
that we have no free will.

 At any rate, one of my aims in this paper is simply to bring
this engaging issue to a wider philosophical audience. However,
I also want to argue that the offer is a much better one than
Bell himself believed. I want to show that we may help ourselves
to the metaphysical advantages---locality and Einsteinian realism---but
save free will. Roughly speaking, the trick is to reinterpret
the same formal possibility in terms of backward causation. Instead
of taking the prior state of the physical system in question
to ``constrain'' the experimenter's choice, we may reasonably
take the latter's choice to affect the prior state of the physical
system. The mathematics thus remains the same as in Bell's proposal,
but we give it a different metaphysical gloss.\footnote{%
Strictly
speaking there are two versions of Bell's proposal. One way of
relaxing the Independence Assumption is to take the required
correlation between hidden variables and future measurement settings
to be established by some common factor in their past. The other
is to take the correlation to be %
{\it %
sui generis}, obtaining simply in virtue of the interaction
between the system and measuring device in question. These two
approaches agree on the ``core'' mathematics involved---on the
nature of the correlation between hidden states and measurement
settings---but disagree about the explanation of this correlation.
It is the latter strategy that I wish to defend here. As I note
in Price (forthcoming), its advantages may have been overlooked
in part because it has not been clearly distinguished from the
former strategy, or because the former strategy has seemed more
plausible, in not countenancing backward causation. Bell seems
to have been aware of both versions, and to have regarded both
as incompatible with free will, but it is doubtful whether he
saw them as clearly distinct. In my view the former strategy
is objectionable on grounds that have nothing to do with free
will, namely that it calls for a vast and all-pervasive substructure
in reality to provide the required ``common causes''. The latter
view in contrast is elegant and economical, as compatible as
need be with free will, and appears to respect a temporal symmetry
that other views ignore.}%

\setlength{\parindent}{12pt}%
\setlength{\parskip}{0pt} The paper is in two main parts. The
first part (sections 1-6) presents what might be called the first-order
case for the backward causation approach. In the interests of
accessibility, I begin with a very informal account of Bell's
Theorem, and the EPR experiment on which it is based. I use a
fictional model to illustrate the basic mathematical conflict
between the kind of predictions made by quantum mechanics and
those that Bell showed to follow from the plausible constraints
of a local realism. Readers already familiar with the Bell's
Theorem-by-parable approach will find nothing new here, and are
encouraged to skim to section 3, where I use the model to explain
how the formal possibility that Bell took to deny free will---the
possibility that I recommend we interpret in terms of backward
causation---provides an elegant resolution of the conflict. Section
4 shows how it avoids traditional objections to backward causation,
section 5 discusses Bell's concerns about free will, and section
6 indicates how the informal discussion of the preceding sections
applies to the real world---i.e., to quantum mechanics. The upshot
seems to be that we may avail ourselves of Bell's route to a
local realism about quantum mechanics, provided that we are prepared
to accept that quantum mechanics reveals the presence of backward
causation in the world.

 At this point in the discussion what is striking is that despite
its evident advantages in quantum mechanics (and the weakness
of the case against it), the backward causation approach still
seems rather unpalatable. The second part of the paper attempts
to diagnose and treat this aversion. Sections 7-13 develop a
kind of second-order case for the approach, intended to clarify
its consequences for physics and hence to show that it is very
much more appealing than it seems at first sight.

 One motivation for this second project is that because the relationship
between causation and physical theory is itself obscure and philosophically
problematic, it is very far from obvious what the backward causation
proposal actually amounts to, in physical terms. Of particular
relevance here is the fact that the temporal asymmetry of causation
is itself rather puzzling, given the predominant symmetry of
physical theory. As we shall see, this puzzle turns out to provide
a useful entry point into the tangle of issues that needs to
be unravelled, in order to clarify the physical content of the
main proposal. In sections 7-9 I defend the view that the asymmetry
of physical dependency---roughly, the fact that the future depends
on the past but not %
{\it vice versa}---is anthropocentric in origin, a kind of projection
of own temporal asymmetry. An attractive feature of this view
is that it dissolves the apparent tension between the predominant
temporal symmetry of physics and the asymmetry of dependency:
since the latter isn't objective, there is no conflict %
{\it in re}, as it were. The cure may seem worse than the disease,
however, and much of the work in this part of the paper will
be devoted to arguing that the view provides a sufficiently close
approximation to objectivity to account for our intuitions concerning
the direction of dependency.

 All the same, the view that the direction of causation is not
genuinely objective may seem rather a mixed blessing for my main
project. It might explain our intuitive resistance to the idea
of backward causation in quantum mechanics, but doesn't it throw
the baby out with the bath water? If the direction of causation
is not objective, what could be the physical content of the claim
that quantum mechanics should be interpreted as revealing that
some causation is ``backward'' rather than ``forward''? Section
10 addresses this important objection, noting that the view that
causal directedness is subjective leaves room for two sorts of
objective correlational structures in the world. From the anthropocentric
viewpoint one sort of structure looks causally ``monotonic'',
whereas the other permits a mixed interpretation, so that the
temporal perspective of the interpreter imposes a dominant but
not a universal causal orientation. This will give us a useful
characterisation of the objective core of what might more accurately
be called the %
{\it advanced action} %
{\it interpretation}. It simply amounts to the suggestion that
the correlational structure of the microworld is of the latter
(non-classical) kind.\footnote{This
enables us to clarify the assertion above that the backward causation
interpretation has the same objective core as Bell's own ``no
free will'' model: both amount to the suggestion that quantum
mechanics shows that what is in world is simply a particular
pattern of correlations (a pattern that classical physics really
had no business to exclude {\it a priori}%
. Whether we choose to interpret this pattern in
terms of predetermination or backward causation thus turns out
to be in an important sense beside the point---the brute physical
facts are the same in either case.} %
More surprisingly, we shall see in section 11 that this characterisation
suggests a powerful symmetry argument in favour of the advanced
action proposal---an argument which has been overlooked, I think,
because it has been obscured by the more familiar asymmetry of
dependency (rightly believed %
{\it not} to conflict with physical symmetry principles). Thus
we shall be led to conclude that the advanced action approach
is not only both physically and metaphysically respectable, but
seems to have a striking advantage, over and above its application
to quantum mechanics.

\setlength{\parindent}{12pt}%
\setlength{\parskip}{0pt} The paper also includes a brief account
of an interesting new class of Bell-like results in quantum mechanics,
which appear to yield similar conclusions to Bell's Theorem by
more direct means. I extend the parable of the earlier sections
to describe these results in informal terms, and to make the
point that they do nothing to weaken the case for advanced action
in quantum mechanics---on the contrary, if anything, since they
simply add new weight to the metaphysical burden from which advanced
action promises to deliver us.

 In case technically inclined readers should be disappointed,
however, I want to make it clear at the outset that I am not
offering the formal details of an interpretation or extension
of quantum mechanics which embodies these ideas. I shall explain
in informal terms how there comes to be a formal possibility
of this kind---how it avoids the challenge to local realism posed
by Bell's Theorem. But my main concern is to show that the strategy
is very much more appealing than Bell took it to be, and therefore
worthy of a great deal more attention from both physicists and
philosophers than it has received so far.\footnote{A few
physicists have been more attracted to the advanced action
approach than Bell himself was. For references to some of their
work, see the concluding footnote.} My thesis is that an exceptionally
promising route to a satisfying
resolution of some of the most profound puzzles in the history
of science lies unexplored and almost unnoticed. The aim of this
paper is to try to clear away some of the conceptual tangles
that have obscured its promise for so long.

\begin{center}%
\setlength{\parindent}{12pt}%
\setlength{\parskip}{0pt}%
{\it 1. The man who proved Einstein was wrong?}%
\footnote{This
description of Bell is due to Gribbin (1990). In this article---a
tribute to Bell following his death in October 1990---Gribbin
gives a useful description of the EPR experiment and Bell's contribution,
and also mentions some of Bell's remarks on the possibility of
abandoning the Independence Assumption. For a comment taking
issue with Gribbin's characterisation of Bell in these terms,
see my (1991a).}%

\end{center}%
\setlength{\parindent}{12pt}%
\setlength{\parskip}{0pt}The most puzzling consequences of quantum
mechanics arise in what are known as the EPR experiments. The
crucial feature of these cases is that they involve a pair of
particles, or physical systems, which interact and then move
apart. Providing the interaction is set up in the right way,
quantum theory shows that the results of measurements on one
particle enable us to predict the results of corresponding measurements
on the other particle. For example we might predict the result
of measuring the position of particle 1 by measuring the position
of particle 2, or predict the result of measuring the momentum
of particle 1 by measuring the momentum of particle 2.

 This was the feature that interested Einstein, Podolsky and
Rosen in the famous (1935) paper in which they first drew attention
to these cases. The paper sought to undermine what was already
becoming the orthodox interpretation of the fact that quantum
theory shows that it is impossible to determine accurately and
simultaneously both the position and the momentum of a physical
system. The orthodoxy---the ``Copenhagen Interpretation'', as
it came to be called---was that quantum systems do not have classical
properties such as position and momentum, except when an appropriate
measurement is made. Einstein wanted to argue that the restriction
on measurement was merely epistemological, however. His motivation
lay in his realism. He disliked the Copenhagen view that the
nature of reality could depend on what humans choose to observe,
and believed that the features of quantum mechanics that Bohr
and others took as evidence of deep entanglement between observation
and reality were really a reflection of the fact that the theory
gives only a partial description of reality. As he saw, the crucial
question is therefore whether the quantum mechanical description
of reality can be considered to be complete. Does it say all
there is to be said about a physical system, or are there further
facts about the physical world not captured by quantum mechanics?

 The two-particle systems seemed to provide the decisive argument
that Einstein was looking for. With Podolsky and Rosen, he argued
that the existence of such systems showed that quantum theory
must indeed be incomplete. For if we can predict either the measured
position or the measured momentum of a particle without interfering
with it in any way, then it must have some property responsible
for the results of those measurements. If we measure the position
of particle 2 we infer that particle 1 has a definite position.
If we measure the momentum of particle 2 we infer that particle
1 has a definite momentum. But since in neither case do we do
anything to particle 1 itself, it must have a definite position
and momentum, regardless of what we do to particle 2. In other
words it must have properties not described by quantum mechanics.
Quantum mechanics must be incomplete.

 The EPR argument failed by and large to sway supporters of the
Copenhagen interpretation, but this is perhaps due more to the
obscurity of the Copenhagen response than to any compelling counter-argument
it brought to light. With the benefit of hindsight we would probably
now say that Einstein was right, had Bell not unearthed a remarkable
sting in the tail of the EPR experiment. Einstein, Podolsky and
Rosen had assumed that what we choose to measure on particle
2 could not have an effect on the distant particle 1. For example,
measuring the position of particle 2 could not somehow ``bring
it about'' that particle 1 had a definite position. In other
words, the EPR argument for the incompleteness of quantum mechanics
assumes that physical effects are %
{\it local}---that there is no action at a distance. The sting
is that other features of the quantum mechanical description
of certain EPR cases seem to show that any more complete (``hidden
variable'') theory would have to be non-local. It would have
to reject the very assumption on which the EPR argument depends.
Einstein's allies thus find themselves in an unfortunate dilemma.
To make a hidden variable theory work---to make it consistent
with quantum mechanics---they have to abandon the assumption
that enabled them to argue from the possibility of the EPR experiment
to the conclusion that there must be such a theory.\footnote{%
It
is important to appreciate that this does not show that Einstein
was wrong: it simply saves his opponents from what would otherwise
be a very serious objection. A common misconception is that Bell's
argument excludes hidden variable (HV) theories {\it tout court}.
It does not. Even leaving aside the loophole discussed
in the present paper, Bell's result counts only against
{\it local} HV views, leaving open the possibility of non-local
HV theory---while non-locality itself can hardly be held to be
a decisive failing in HV views if other views need it as well,
as Bell's result suggests that they do.}

\setlength{\parindent}{12pt}%
\setlength{\parskip}{0pt} The sting turns on the predictions
that quantum theory makes about the correlations between the
results of the various possible measurements on the two particles
of certain EPR systems. It was the significance of these correlations
that was first noticed by Bell in the mid-1960s. Bell considered
a variant of the original EPR case (a version originally described
by Bohm, (1951), pp.\ 614-19). Fortunately for lay readers, we
don't need to know the details of Bohm's case or Bell's Theorem
to appreciate its puzzling character. The essential features
can be described in terms of an informal and much more commonplace
analogue. As we shall see, very little mathematical thought is
required to show that if analogous correlations were to arise
in familiar regions of the world, they would strike us as very
odd indeed.%
\footnote{%
I
have adapted the following account from those given in several
lucid and entertaining papers by the physicist N. David Mermin---see
particularly his (1985) and (1981).
}

\begin{center}%
\setlength{\parindent}{12pt}%
\setlength{\parskip}{12pt}%
{\it 2. Ypiaria.}%
\end{center}%
\setlength{\parskip}{0pt}By modern standards the criminal code
of Ypiaria%
\footnote{%
Pronounced,
of course, ``E-P-Aria''.%
} allowed its police force excessive powers of arrest and interrogation.
Random detention and questioning were accepted weapons in the
fight against serious crime. This is not to say that the police
had an entirely free hand, however. On the contrary, there were
strict constraints on the questions the police could address
to anyone detained in this way. One question only could be asked,
to be chosen at random from a list of three: (1) Are you a murderer?
(2) Are you a thief? (3)~Have you committed adultery? Detainees
who answered ``yes'' to the chosen question were punished accordingly,
while those who answered ``no'' were immediately released. (Lying
seems to have been frowned on, but no doubt was not unknown.)

\setlength{\parindent}{12pt}%
\setlength{\parskip}{0pt} To ensure that these guide-lines were
strictly adhered to, records were required to be kept of every
such interrogation. Some of these records have survived, and
therein lies our present concern. The records came to be analysed
by the psychologist Alexander Graham Doppelg\"{a}nger, known
for his work on twins and co-operative behaviour. Doppelg\"{a}nger
realised that amongst the many millions of cases in the surviving
records there were likely to be some in which the Ypiarian police
had interrogated both members of a pair of twins. He was interested
in whether in such cases any correlation could be observed between
the answers given by each twin.

 As we now know, Doppelg\"{a}nger's interest was richly rewarded.
He uncovered the two striking and seemingly incompatible correlations
now known collectively as %
{\it Doppelg\"{a}nger's Twin Paradox}. He found that

\begin{description}%

\item[{\sc same}] %
When each member of a pair of twins was asked
the same question, both always gave the same answer;
\end{description}%
and that
\begin{description}%

\item[{\sc diff}] When each member of a pair of twins was asked
a different question, they gave the same answer on close to 25\%
of such occasions.

\end{description}%
\setlength{\parindent}{12pt}It may not be immediately
apparent that these results are in any way incompatible. But
Doppelg\"{a}nger reasoned as follows. %
{\sc same} means that whatever it is that disposes Ypiarians
to answer Y or N to each of the three possible questions 1, 2
and 3, it is a disposition that twins always have in common.
For example, if ``YYN'' signifies the property of being disposed
to answer Y to questions 1 and 2 and N to question 3, then correlation

 {\sc same} implies that if one twin is YYN then so is
his or her sibling. Similarly for the seven other possible such
states: in all, for the eight possible permutations of two possible
answers to three possible questions. (The possibilities are the
two ``homogeneous'' states YYY and NNN, and the six ``inhomogeneous''
states YYN, YNY, NYY, YNN, NYN and NNY.)

 Turning now to %
{\sc diff}, Doppelg\"{a}nger saw that there were six
ways to ask a different question to each of a pair of twins:
the possibilities we may represent by Q%
{\normalsize $_{12}$}, Q%
{\normalsize $_{21}$}, Q%
{\normalsize $_{13}$}, Q%
{\normalsize $_{31}$}, Q%
{\normalsize $_{23}$} and Q%
{\normalsize $_{32}$} (``Q%
{\it %
{\normalsize $_{ij}$}}'' thus signifies that the first twin is
asked question %
{\it i} and the second question %
{\it j}). How many of these possibilities would produce the
same answer from both twins? Clearly it depends on the twins'
shared dispositions. If both twins are YYN, for example, then
Q%
{\normalsize $_{12}$} and Q%
{\normalsize $_{21}$} will produce the same response (in this
case, Y) and the other four possibilities will produce different
responses. So if YYN twins were questioned at random, we should
expect the same response from each in about $1/3$ of all cases.
Similarly for YNY twins, YNN twins, or for any of the other inhomogeneous
states. And for homogeneous states, of course, all six possible
question pairs produce the same result: YYY twins will always
answer Y and NNN twins will always answer N.

 Hence, Doppelg\"{a}nger realised, we should expect a certain
minimum correlation in these different question cases. We cannot
tell how many pairs of Ypiarian twins were in each of the eight
possible states, but we can say that whatever their distribution,
confessions should correlate with confessions and denials with
denials in at least $1/3$ of the different question interrogations.
For the figure should be $1/3$ if all twins are in inhomogeneous
states, and higher if some are in homogeneous states. And yet,
as %
{\sc diff} describes, the records show a much lower
figure.

 Doppelg\"{a}nger initially suspected that this difference might
be a mere statistical fluctuation. As newly examined cases continued
to confirm the same pattern, however, he realised that the chances
of such a variation were infinitesimal. His next thought was
therefore that the Ypiarian twins must generally have known what
question the other was being asked, and determined their own
answer partly on this basis. He saw that it would be easy to
explain %
{\sc diff} if the nature of one's twin's question could
influence one's own answer. Indeed, it would be easy to make
a total anti-correlation in the different question cases be compatible
with %
{\sc  same}---with total correlation in the same question
cases.

 Doppelg\"{a}nger investigated this possibility with some care.
He found however that twins were always interrogated separately
and in isolation. As required, their chosen questions were selected
at random, and only after they had been separated from one another.
There therefore seemed no way in which twins could conspire to
produce the results described in %
{\sc same} and %
{\sc diff}. Moreover, there seemed a compelling physical
reason to discount the view that the question asked of one twin
might influence the answers given by another. This was that the
separation of such interrogations was usually space-like in the
sense of special relativity; in other words, neither interrogation
occurred in either the past or the future light-cone of the other.
(It is not that the Ypiarian police force was given to space
travel, but that light travelled more slowly in those days. The
speed of a modern carrier pigeon is the best current estimate.)
Hence according to the principle of the relativity of simultaneity,
there was no determinate sense in which one interrogation took
place before the other. How then could it be a determinate matter
whether interrogation 1 influenced interrogation 2, or vice versa?%
\footnote{%
There
are a number of concerns that might arise here, but the one that
Doppelg\"{a}nger seems to have found most pressing is this: if
we allow space-like influences of this kind, then if it is not
to be an arbitrary matter at what time a given influence ``arrives'',
certain inertial frames must be physically distinguished from
others, in violation of the spirit of special relativity.}

\setlength{\parindent}{12pt}%
\setlength{\parskip}{0pt} How are we to explain Doppel\-g\"{a}nger's
remarkable observations? Doppel\-g\"{a}ng\-er himself seems reluctantly
to have favoured the telepathic hypothesis---the view that despite
the lack of any evident mechanism, and despite the seeming incompatibility
with the conceptual framework of special relativity, Ypiarian
twins were capable of being ``instantaneously'' influenced by
their sibling's distant experiences. Doppelg\"{a}nger was well
aware that there is an hypothesis that explains %
{\sc same} and %
{\sc diff} without conflicting with special relativity.
It is that the twins possess not telepathy but precognition,
and thus know in advance what questions they are to be asked.
However, he felt that this interpretation would force us to the
conclusion that the Ypiarian police interrogators were mere automatons,
not genuinely free to choose what questions to ask their prisoners.
Other commentators have dismissed the interpretation on different
grounds, claiming that it would give rise to causal paradoxes.

 In my view neither of these objections is philosophically well-founded.
The relativity-friendly alternative that Doppelg\"{a}nger rejects
is certainly counterintuitive, but it is not absurd. Given the
nature of the case, any workable explanation will be initially
counterintuitive. What matters is whether that intuition withstands
rigorous scrutiny, and of course how much gain we get for any
remaining intuitive pain. Doppelg\"{a}nger himself was well aware
of the gains that would flow from the interpretation in question
(especially that it saves special relativity), but thought the
pain too high. I want to show that he was mistaken.\footnote{%
In
saying this I am claiming no special insight or competence with
respect to Ypiarian psychology, of course. My suggestion is simply
that the specialists in that field have been led astray by considerations
which fall more within philosophy than within their own discipline.
}

\begin{center}%
\setlength{\parindent}{12pt}%
\setlength{\parskip}{12pt}%
{\it 3. Advanced action: how it explains the Twin Paradox.}%
{\normalsize }

\end{center}%
\setlength{\parskip}{0pt}Very little is known about the factors
which must have governed an Ypiarian's answers to the three questions
permitted under Ypiarian law. The surviving records tell us what
was said but not in general why it was said. The psychological
variables are hidden from us, and must be inferred, if at all,
from the behavioural data to which we have access. However, the
puzzling character of the data is easily explained if we allow
that the relevant variables display what physicists call advanced
action: the property that the underlying psychological variables
may depend on the %
{\it future} experiences of the agents concerned. Specifically,
what is required is that pairs of twins who are in fact going
to be asked different questions are therefore more likely to
adopt an inhomogeneous state that yields different answers to
those questions. Hidden variables are thus dependent on the %
{\it fate} of the agents concerned, as well on their %
{\it history}.

 Consider for example a pair of twins %
{\it T}%
{\it %
{\normalsize $_{dum}$}} and %
{\it T}%
{\it %
{\normalsize $_{dee}$}} %
whose fate is in fact to be asked questions 2 and
3 respectively. Doppelg\"{a}nger's correlation %
{\sc %
{\normalsize diff}} suggests that this fate has the effect of
making it less likely that %
{\it T}%
{\it %
{\normalsize $_{dum}$}} and %
{\it T}%
{\it %
{\normalsize $_{dee}$}} %
{\sc $_{ }$}will be in the states YYY, YNN, NYY and NNN that
yield the same answer to questions 2 and 3; and correspondingly
more likely that %
{\it T}%
{\it %
{\normalsize $_{dum}$}} and %
{\it T}%
{\it %
{\normalsize $_{dee}$}} %
{\it %
{\sc $_{ }$}}will be in one of the states YYN, YNY, NYN and NNY
that yield different answers to these questions.

 Unlike Doppelg\"{a}nger's instantaneous action-at-a-distance,
this advanced action proposal does not conflict with special
relativity. This is because the point at which twins become coupled---whether
their conception, their birth, or some later meeting---lies well
within the light-cones of both their later interrogations. The
effect is not instantaneous and not at a space-like distance.
And it needs no mysterious carrier. It has the twins themselves,
who bear the marks of their future as they bear the marks of
their past.

 Why has this interpretation not appealed to Doppelg\"{a}nger
and other specialists in this field? As noted earlier, it is
because they think that there is something absurd or contradictory
in the idea that present events might exhibit this kind of correlation
with future events. The objection resolves into two main strands.
The first, which seems to have been the more influential in Doppelg\"{a}nger's
own case, is the intuition that advanced action leads to fatalism---that
it is incompatible with the ordinary supposition that the future
events concerned are ones with respect to which we (or the Ypiarian
interrogators) have independent present control and freedom of
choice. We shall return to this strand of the objection in section
5.

 The second strand of the objection is the claim that the correlations
required for advanced action could be exploited to yield ``causal
paradoxes'' of one kind or another. This claim depends on a venerable
line of reasoning, essentially the so-called ``bilking argument''.
But what are these paradoxes, and need they arise in the Ypiarian
case? I think we can show that they might not.

\begin{center}%
\setlength{\parskip}{12pt}%
{\it 4. Causal paradox?}%
{\normalsize }

\end{center}%
\setlength{\parskip}{0pt}The supposed paradox is familiar from
countless science fiction stories. The hero travels into her
own past and takes steps to ensure that she will not exist at
the time she left---she kills her young self, introduces her
mother-to-be to contraception, persuades her Author to take up
accounting, or something of the kind. The results are contradictory:
the story tells us that something is both true and not true.
It is like being told that it was the best of times and the worst
of times, except that we are offered an account of how this contradictory
state of affairs came to pass. This has the literary advantage
of separating the beginning of the story from the end, and the
logical advantage of allowing us to conclude that something in
the offered account is false. In physical rather than literary
mode we can thus argue against the possibility of time travel:
if there were time travel we could design an experiment with
contradictory results; %
{\it ergo}, there is no time travel.\footnote{I
am simplifying here, of course. For one thing it is clear that
even given the hypothesis of time travel, we are never actually
justified in expecting the experiment to yield contradictory
results, for logic alone rules that out. A number of authors
have made this the basis of a defence of the possibility of time
travel against the bilking argument. (See Horwich (1975), Lewis
(1976) and Thom (1974), for example.) This issue is not directly
relevant to our present concerns, which exploit a much larger
loophole in the bilking argument. In passing, however, let me
record my view---similar to that of Horwich (1987), ch. 7---that
the bilking argument survives the former challenge. Roughly speaking,
it shows us that hypothesis of time travel can be made to imply
propositions of arbitrarily low probability. This is not a classical
{\it reductio}%
, but it is as close as science ever gets.%
} And as for time travel, so for backward causation. Causing
one's young self to have been pushed under a bus is just as suicidal
as travelling in time to do the deed oneself.

\setlength{\parindent}{12pt}%
\setlength{\parskip}{0pt} In summary, then, the view that backward
causation leads to causal paradoxes rests on the claim that if
there were such causation, it could be exploited to allow retroactive
suicide and other less dramatic but equally absurd physical results.
But is this true of the kind of backward influence I claimed
to find in the Ypiarian case? Could YES, the Ypiarian Euthanasia
Society,\footnote{``YES for the right to say NO!'', as their slogan goes.} %
exploit it in the interests of painless deaths for its aging
members?

\setlength{\parindent}{12pt}%
\setlength{\parskip}{0pt} Unfortunately not, I think. For what
is the earlier effect---the claimed result of the later fact
that a pair of twins %
{\it T}%
{\it %
{\normalsize $_{dum}$}} and %
{\it T}%
{\it %
{\normalsize $_{dee}$}} %
{\sc $_{ }$}are asked questions 2 and 3, say? It is that the
state in which %
{\it T}%
{\it %
{\normalsize $_{dum}$}} and %
{\it T}%
{\it %
{\normalsize $_{dee}$}} %
{\sc $_{ }$}separate is less likely to be one of those (YYY,
YNN, NYY or NNN) in which questions 2 and 3 give the same response.
For simplicity let us ignore the probability, making the effect
correspondingly stronger. Assume, in other words, that such a
fate %
{\it guarantees} that %
{\it T}%
{\it %
{\normalsize $_{dum}$}} and %
{\it T}%
{\it %
{\normalsize $_{dee}$}} %
{\sc $_{ }$}will not be in one of these states. Retroactive suicide
then requires that this effect be wired to produce the desired
result---that a device be constructed that kills the intended
victim if %
{\it T}%
{\it %
{\normalsize $_{dum}$}} and %
{\it T}%
{\it %
{\normalsize $_{dee}$}} %
{\sc $_{ }$}do not have one of the excluded states. (This machine
might be too generous, if other future possibilities do not prevent
the past state of affairs that triggers it; but this won't worry
the members of YES, who will not be concerned that they might
already have killed themselves---accidentally, as it were.) The
trigger thus needs to detect the relevant states of %
{\it T}%
{\it %
{\normalsize $_{dum}$}} and %
{\it T}%
{\it %
{\normalsize $_{dee}$}}, %
{\it before the occurrence of the claimed future influence on
these states}---before %
{\it T}%
{\it %
{\normalsize $_{dum}$}} and %
{\it T}%
{\it %
{\normalsize $_{dee}$}} are next interrogated by the Ypiarian
police. But why suppose it is physically possible to construct
such a detector? We know these states can be detected by the
process of interrogation---in effect they are dispositions to
respond to such interrogation in a certain way---but this is
not to say that they are ever revealed in any other way.

 We thus have the prospect of an answer to the bilking objection.
The thought experiments involved rest on an assumption that might
simply be false. The supposedly paradoxical experiment might
be physically impossible. This is a familiar kind of response
to such arguments in science. Consider for example the old argument
that space must be infinite, since if it were finite one could
journey to the edge and extend one's arm. One response to this
is to point out that even if space were finite the required journey
might be physically impossible, because a finite space need have
no edges.

 It is perhaps impossible at this distance to adjudicate on the
Ypiarian case. Doppelg\"{a}nger's work notwithstanding, we know
too little of Ypiarian psychology to be able to say whether the
relevant states would have been detectable before interrogation.
It is enough that because the bilking argument depends on this
assumption, the backward causation proposal remains a live option.\footnote{The
fact that the bilking argument depends on an assumption of this
kind was pointed out by Michael Dummett (1964). Later we shall
see that quantum mechanics is tailor-made to exploit Dummett's
loophole.
}

\begin{center}%
\setlength{\parindent}{12pt}%
\setlength{\parskip}{12pt}%
{\it 5. The fatalist objection.}%
{\normalsize }

\end{center}%
\setlength{\parskip}{0pt}Now to Doppelg\"{a}nger's own main
concern about the advanced action interpretation, namely that
it seems to deny free will. If the Ypiarian interrogators' choices
of questions are correlated with the earlier psychological states
of the twins concerned, then surely their apparent freedom of
choice is illusory. For when they come to ``decide'' what questions
to ask their ``choice'' is already fixed---already ``written''
in the mental states of their interviewees. This is the %
{\it fatalist} objection to advanced action.

The first thing to note about the fatalist objection is that
it tends to slide back into a version of the bilking argument.
If we think of ``already determined'' as implying ``accessible'',
then we seem to have the basis of a paradox-generating thought
experiment. Since we have already discussed the bilking argument,
let us assume that ``already determined'' does not imply ``accessible''.
What then remains of the fatalist point?

The strategic difficulty now is to set aside this objection
to advanced action without calling on a philosophical dissertation
on the topic of free will. Two points seem to be in order. The
first is that even if the argument were sound, it would not show
that advanced action is physically impossible. Rather it would
show that advanced action is physically incompatible with the
existence of free will. In the interests of theoretical simplicity
the appropriate conclusion might then be that there is no free
will. Free will might then seem another piece of conceptual anthropocentrism,
incompatible (as it turns out) with our best theories of the
physical world. So the fatalist objection is strictly weaker
than the causal paradox argument. If successful the latter shows
that advanced action is physically impossible; the former, at
best, merely that it is incompatible with free will.

 The second thing to be said about the fatalist
objection is that it has a much more familiar twin. This is the
argument that if statements about our future actions are ``already''
either true or false (even if we can't know which), then we are
not free to decide one way or the other. There are differing
views of this argument among philosophers. The majority opinion
seems to be that it is fallacious, but some think that the argument
does rule out free will, and others that free will is only saved
by denying truth values to statements about the future. On the
last point, physicists perhaps have more reason than most to
grant determinate truth values to statements about the future.
Contemporary physics is usually held to favour the four dimensional
or ``block universe'' view of temporal metaphysics, whereas the
thesis that future truth values are indeterminate is typically
thought to require the rival ``tensed'' view of temporal reality.

 If we accept that statements about the future have truth values
then there are two possibilities. If the ordinary fatalist argument
is sound, then there is no free will. But in this case the fatalist
argument against advanced action is beside the point. If there
is no free will anyway, then it is no objection to advanced action
if it implies that there is none. If the ordinary argument is
unsound, on the other hand, then so surely is the backward version.
For the arguments are formally parallel. If the refutation of
the forward version does not depend on what makes it the forward
version---in effect, on the special status of future tensed statements---then
it depends on what the two versions have in common. If one argument
fails then so does the other.

 Indeed, what it seems plausible to say in the future case is
something like this: statements about the future do have truth
values, and some of these statements concern events or states
of affairs which do stand in a relation of constraint, or dependence,
with respect to certain of our present actions. However, what
gives direction to the relation---what makes it appropriate to
say that it is our actions that ``fix'' the remote events, rather
than vice versa---is that the actions concerned %
{\it are} our actions, or products of our free choices. The
fatalist's basic mistake is to fail to notice the degree to which
our talk of (directed) dependency rides on the back of our conception
of ourselves as free agents. Once noted, however, the point applies
as much in reverse, in the special circumstances in which the
bilking argument is blocked.

At any rate, the more basic point is that the fatalist
argument usefully distinguishes the backward case only if two
propositions hold: (a) the classical (future-directed) fatalist
argument is valid, and thus sound if future tensed statements
do have truth values; and (b) there is a significant difference
between the past and the future, in that past tensed statements
do have truth values, whereas future tensed statements do not.
Proposition (a) is a rather unpopular position in a philosophical
debate of great antiquity. Proposition (b) is perhaps a less
uncommon position, but one that modern physicists seem to have
more reason than most to reject. Taken together, as they need
to be if the fatalist objection to the advanced action interpretation
is not to collapse, these propositions thus form an unhealthy
foundation for an objection to an otherwise promising scientific
theory.

\begin{center}%
\setlength{\parskip}{12pt}%
{\it 6. Quantum mechanics and backward causation.}%
{\sc %
{\normalsize }}

\end{center}%
\setlength{\parskip}{0pt}The point of the Ypiarian example lies,
of course, in the fact that it exactly mirrors the puzzling behaviour
of certain two-particle quantum-mechanical systems. Doppelg\"{a}nger's
{\sc same} is effectively the feature of EPR systems
on which the original EPR argument relied. And %
{\sc diff} is the additional feature whose conflict
with %
{\sc same} was noted by Bell in 1965. The case mirrors
Bohm's version of the EPR experiment. The pairs of twins correspond
to pairs of spin-$1/2$ particles in the singlet state. The act
of asking a twin one of three specified questions corresponds
to the Stern-Gerlach determination of the spin of such a particle
in one of three equi-spaced directions perpendicular to the line
of flight. The answers Y and N correspond on one side to the
results ``spin up'' and ``spin down'' and on the other side to
the reverse. Thus a case in which both twins give the same answer
corresponds to one in which spin measurements give opposite results.
Correlations %
{\sc same} and %
{\sc diff} follow from the following consequence of
quantum mechanics: when the orientation of the Stern-Gerlach
measurements differ by an angle a then the probability of spin
measurements on each particle yielding opposite values is $cos$%
{\normalsize $^{2}$}($\alpha/2$). This probability is 1 when $\alpha=0$ (%
{\sc same}) and $1/4$ when $\alpha=\pm120^\circ$ (%
{\sc diff}).

 These predictions have been confirmed in a series of increasingly
sophisticated experiments.\footnote{Most
notably those of Alain Aspect %
{\it et. al.}---see Mermin (1985), pp.\ 45-6.} %
 Thus if you thought the proper response to the Ypiaria story
was that it was simply unrealistic, you should think again. Not
only is the perplexing behaviour of Ypiarian twins theoretically
and practically mirrored in quantum mechanics, but quantum mechanics
actually tells us what sort of neurophysiology would make people
behave like that. All we have to suppose is that the brains of
Ypiarian twins contain the appropriate sort of correlated spin-$1/2$
 particles (one particle in each twin), and that interrogation
causes a spin determination, the result of which governs the
answer given.

\setlength{\parindent}{12pt}%
\setlength{\parskip}{0pt} As in the Ypiarian case, it is easy
to explain Bell's results if we allow that particle states can
be influenced by their fate as well as their history. One way
to do it is to allow for a probabilistic ``discoupling factor''
which depends on the actual spin measurements to be performed
on each particle and which influences the underlying spin properties
of the particles concerned. We simply say that the production
of such particle pairs is governed by the following constraint:

\begin{description}%
\item [{\sc set}] In those directions G and H (if any) in which the spins
are going to be measured, the probability that the particles
have opposite spins is $cos$%
{\normalsize $^{2}$}$(\alpha/2)$, where $\alpha$ is the angle between G and
H.

\end{description}%

\setlength{\parindent}{12pt}Note that this condition
refers to the %
{\it fate} of the particles concerned; it allows that their
present properties are in part determined by the %
{\it future} conditions they are to encounter. Thus it explicitly
violates Bell's Independence Assumption.\footnote{{\sc set} should be taken
simply as an illustration of the
general strategy, of course. Recall that we began with Bell's
own observation that if we are prepared to abandon the Independence
Assumption, his results no longer stand in the way of a local
hidden variable theory for quantum mechanics---a local realist
theory, in Einstein's sense. The issue of the best form for such
a theory is a technical matter, beyond the scope of the present
paper. The present argument is simply that %
{\it in philosophical terms} %
 Bell's loophole is much more attractive than it
has usually been taken to be, and hence that the technical strategy
it embodies has been unjustly neglected.
}

\setlength{\parindent}{12pt}%
\setlength{\parskip}{0pt} How does %
{\sc set} cope with the kind of objections to advanced action
that we dealt with in the Ypiarian case? We saw that the causal
paradox objection rested on assumption that the claimed earlier
effect could be detected in time to prevent the occurrence of
its supposed later cause. What does this assumption amount to
in the quantum mechanics case? Here the claimed earlier effect
is the arrangement of spins in the directions G and H which are
later to be measured. But what would it take to detect this arrangement
in any particular case? It would take, clearly, a measurement
of the spins of particles concerned in the directions G and H.
However, such a measurement is precisely the kind of event which
is being claimed to have this earlier effect. So there seems
to be no way to set up the experiment whose contradictory results
would constitute a causal paradox. By the time the earlier effect
has been detected its later cause has already taken place.\footnote{In
effect, quantum mechanics thus builds in exactly what we need
to exploit the Dummettian loophole in the bilking argument.
}

\setlength{\parindent}{12pt}%
\setlength{\parskip}{0pt} The advanced action explanation of
Bell's results thus lacks the major handicap with which it has
usually been saddled. On the other side, it has the advantages
we noted in the Ypiarian case. For one thing, it does not seem
to call for any new field or bearer of the influence that one
measurement exerts on another. If we think of the fate of a particle
as a property of that particle---a property which has a bearing
on the results of its interaction with its twin---then the particles
themselves ``convey'' the relevant influence to its common effect
at the point at which they separate. More importantly, by thus
confining the retro-influence of future measurements to their
past light-cones, the advanced action account avoids action at
a distance, and hence the threat of conflict with special relativity.

 The extent of the last advantage clearly depends on how much
alternative explanations do conflict with special relativity.
This has been a topic of considerable discussion in recent
years.\footnote{There
is an up-to-date analysis in Butterfield (1994), and a survey
of the field in Maudlin (1994), chs. 4-6.
} It is widely agreed that the Bell correlations do not permit
faster than light signalling, but the issue of ``causal influence''
is less straightforward. Whatever the nature of the influence,
the concern seems in part to be that special relativity implies
that any space-like influence will be a backward influence from
the point of view of some inertial frame of reference. Backward
influence has seemed problematic on the grounds we have already
examined: ``It introduces great problems, paradoxes of causality
and so on.'' (Bell again, in Davies and Brown 1986, p.\ 50) Now
if this were the only problem with Bell's preferred non-local
influences, that fact would weaken our case for preferring backward
to space-like influence; for it would mean that in rejecting
the usual argument against backward causation we would also be
removing the main obstacle to Bell's own faster-than-light interpretation
of the phenomena.

\setlength{\parindent}{12pt}%
\setlength{\parskip}{0pt} There is another problem with Bell's
view, however, namely the concern we noted in the Ypiarian case:
a space-like influence seems to distinguish one inertial frame
of reference from all the rest---it picks out the unique frame
according to which the influence concerned is instantaneous.
This appears to contradict the accepted interpretation of special
relativity, namely that all inertial frames are physically equivalent.
It was this consequence to which Bell was referring in the passages
I quoted at the beginning of the paper. As he puts it elsewhere,
the view commits us to saying that ``there %
{\it are} influences going faster than light, even if we cannot
control them for practical telegraphy. Einstein local causality
fails, and we must live with this.'' (1987, p.\ 110) As I have
emphasised, the advanced action interpretation requires no such
reconstruction of special relativity. So unless there is more
to be held against this approach than the charges of fatalism
and causal paradox, it would appear to offer a rather promising
explanation of the Bell correlations.

 I have concentrated on the EPR cases because it is here that
in virtue of the threatened conflict with special relativity
the advantages of backward influence (or advanced action) are
most apparent. Concerning quantum mechanics more generally, I
think its appeal turns on its ability to revitalise that other
aspect of Einstein's world view, namely the conviction that quantum
mechanics provides an incomplete description of the physical
world. As I noted earlier, the extent to which Bell's Theorem
undermines this Einsteinian view is commonly exaggerated. The
present loophole aside, Bell's result simply undermines %
{\it local} versions of such a view---which, given that it appears
to undermine locality generally, can hardly be counted a decisive
objection to Einstein. All the same, the present loophole does
seem to swing the argument strongly in favour of the Einstein
view: given that orthodox quantum mechanics does not embody the
required ``backward influence'', only a model which takes the
orthodox theory to be incomplete will be capable of doing so.
Only some version of the Einstein view seems able to save locality
in this way.\footnote{Far
from being the man who proved Einstein wrong, Bell thus appears
in this light to be the man who provided a new reason to think
that Einstein was right.
}

\setlength{\parindent}{12pt}%
\setlength{\parskip}{0pt} A revitalised Einsteinian view would
have the attractions it always held for the interpretation of
quantum mechanics. Its great virtue is that because it denies
that the collapse of the wave function corresponds to a real
change in the physical system concerned, it does not encounter
the so-called ``Measurement Problem'', which is essentially the
problem of providing a principled answer to the question as to
exactly when such changes take place. The fact that the Measurement
Problem is an artefact of a particular way of interpreting quantum
mechanics has tended to be forgotten as the views of Einstein's
Copenhagen opponents have become the orthodoxy, but it was well
appreciated in the early days of the theory. As Lockwood (1989,
pp.\ 196-7) makes clear, the point of Schr\"{o}dinger's original
use of his famous feline %
{\it gedankenexperiment} was precisely to distinguish these
two ways of looking at quantum theory, and to point out that
unlike the view of the theory that he himself shared with Einstein,
that of their Copenhagen opponents would be saddled with the
problem of the nature and timing of measurement.

 Although it avoids the standard Measurement Problem, it might
seem that the view I am suggesting---Einsteinian realism with
advanced action---will face a measurement problem of its own.
For if the claim is that earlier hidden variables are affected
by later measurement settings, don't we still need a principled
account of what counts as a measurement? This is a good point,
but the appropriate response seems not to be to try to distinguish
measurements from physical interactions in general, but to note
that it is a constraint on any satisfactory development of this
strategy for quantum theory that the advanced effects it envisages
be products of physical interactions in general, rather than
products of some special class of ``measurement'' interactions.

 This point is relevant to another charge that might be levelled
against the suggested approach. It might be argued that the approach
fails to respect the core thesis of Einstein's realism, namely
the principle that there is an objective world whose nature and
reality are independent of human observers. If our present measurements
affect the prior state of what we observe, then surely the external
world is not independent in Einstein's sense. Had we chosen to
make a different measurement, the external world would have been
different. Isn't this very much like the observer-dependence
that Einstein found objectionable in the Copenhagen view?

 I think it is best to answer this charge indirectly. First of
all, I think we may assume that Einstein would not have felt
that his brand of realism was threatened by the observation that
human activities affect the state of the external world in all
sorts of ordinary ways. The existence of trains, planes and laser
beams does not impugn realism. The processes that produce such
things are simply physical processes, albeit physical processes
of rather specialised kinds. Secondly, it seems fair to assume
that backward causation is not in itself contrary to spirit of
Einsteinian realism. On the contrary, a realist of Einstein's
empiricist inclinations might well think that the direction of
causation is a matter to be discovered by physics.\footnote{To
the extent that causation itself is regarded as a physically
respectable notion, at any rate. Other views are possible, of
course, but then the objection to backward causation will not
be specifically that it threatens classical realism.
} But then why should the existence of backward effects of human
activities be any more problematic for realism than the existence
of their ``forward'' cousins? Provided we make it clear that
in the first place the view is that certain %
{\it physical} interactions have earlier effects, and not that
certain specifically %
{\it human} activities do so, the position does not seem to
be one that an empiricist of Einstein's realist persuasions should
object to %
{\it a priori}. For what the view proposes, roughly speaking,
is simply to extend to the case of the past a form of lack of
independence that realists find unproblematic in the case of
the future. The proposal might perhaps be objectionable for other
reasons, but it does not conflict with realist intuitions of
Einstein's sort.

\setlength{\parindent}{12pt}%
\setlength{\parskip}{0pt} In summary, then, the advanced action
approach promises the usual virtues of the ``incompleteness''
interpretation of quantum mechanics favoured by Schr\"{o}d\-inger,
Born, Einstein, Bohm and Bell himself. Unlike other versions
of this view (and apparently %
{\it all} versions of the opposing orthodoxy), it also promises
to save locality, and therefore to avoid the threatened conflict
between quantum mechanics and special relativity. The restrictions
that quantum mechanics places on measurement enable the approach
to exploit a well-recognised loophole in the bilking argument---while
interpreting the required failure of the Independence Assumption
in terms of backward causation seems to sidestep Bell's own concerns
about free will.

\begin{center}%
\setlength{\parskip}{12pt}%
{\it 7. Conceptual inertia and the puzzle of causal asymmetry.}%
{\sc %
{\normalsize }}

\end{center}%
\setlength{\parskip}{0pt}Striking as it is, however, I have
come to appreciate that the above argument has the following
rather disappointing characteristic: it is almost completely
unpersuasive.\footnote{Perhaps
I'm being unduly pessimistic here, but try the following experiment:
imagine that you find the steps in the argument individually
convincing, so that you now accept that the advanced action interpretation
has the advantages described above. Are you now disposed to believe
that quantum mechanics shows that the world contains backward
causation? I suspect not, for the kind of reasons outlined below.
} Even its most receptive audiences seem to find themselves in
the grip of a kind of conceptual inertia---a phenomenon to which,
as it stands, the argument simply makes no concessions. In effect,
the argument assumes that the ordinary view that there is no
backward causation is just another scientific hypothesis, amenable
to revision on empirical grounds. But causation is a pre-scientific
notion, deeply ingrained in folk usage and philosophically problematic.
At the very least, therefore, an argument with such radical conclusions
on the subject needs to be backed up with a kind of metaphysical
user's manual---a guide to some of the philosophical issues in
the background, so that interested parties might have a better
sense as to what they are being invited to endorse.

\setlength{\parindent}{12pt}%
\setlength{\parskip}{0pt} One function for such a guide might
well be cognitive therapy. The unpopularity of the backward causation
approach to the interpretation of quantum mechanics---a field
not noted for metaphysical conservatism!---itself suggests that
the intuitions that oppose it are very deeply ingrained. In order
to give the approach a fair hearing, then, it seems appropriate
to ask where these intuitions come from, and why they are so
resistant to arguments of the above kind. As in psychoanalysis,
uncovering the aetiology of our aversions might be the first
step to their successful treatment.

 Whatever the motivation for the project, however, an obvious
starting point is a reflection on the nature and origins of causal
asymmetry in ``classical'' cases. One of the puzzling features
of our intuitions concerning the temporal asymmetry of causation
and physical dependency is that they seem to be quite independent
of what we know of the temporal symmetry of the underlying physical
processes concerned. This is especially striking in simple microphysical
examples. Consider, for example, a photon emitted from an excited
atom in some distant galaxy, and eventually absorbed in the reverse
process on Earth (perhaps in a detector connected to a telescope).
Suppose that the photon happens to pass through polarisers at
both ends of its journey. By our intuitive standards, nothing
could be more natural than the thought that while the state of
the photon as it nears the Earth might depend on the setting
of the distant polariser (perhaps million of years in its past),
it doesn't depend on the setting of the local polariser (perhaps
only microseconds in its future). Yet if the physics is symmetric---if
from the physical point of view it is simply a conventional matter
which extremity of the photon's journey we choose to regard as
the beginning and which the end---then what could possibly explain
this difference?

 As this example illustrates, there is a %
{\it prima facie} tension between the striking temporal symmetry
of contemporary physical theory and the prevailing intuition
that causation is not only asymmetric but ``unidirectional''---always
oriented in the same temporal direction. True, this tension is
commonly said to be defused by the recognition that causal direction
is a physically contingent feature of the world---``fact-like''
rather than ``law-like'', as physicists often say. But given
that this response usually associates causal direction with the
thermodynamic asymmetry, itself normally taken to be statistical
and thus macroscopic in origin, it is doubtful whether it provides
much support for applying our ordinary causal intuitions
 to microphysics.\footnote{I develop this objection in Price (1992).}

\setlength{\parindent}{12pt}%
\setlength{\parskip}{0pt} In my view the most plausible resolution
of the apparent tension rests on the thesis that the asymmetry
of causal dependency is anthropocentric in origin. Roughly, the
suggestion is that we ourselves are asymmetric in time, and that
this subjective asymmetry comes to be embedded in the way we
talk about the world. To argue this way is not to deny that there
are objective temporal asymmetries in the physical world. It
is simply to point out that some of the apparently ``external''
asymmetries may be ``internal'' in origin.\footnote{The
claim is not undermined by the requirement that the ``internal''
asymmetries concerned should themselves be explicable ``externally''---i.e.,
in terms of objective physical asymmetries in the world---so
long as the latter asymmetries are not those being held to be
projections of our own internal asymmetry. A plausible hypothesis
is that our own existence and temporal asymmetry is ultimately
explicable in terms of the thermodynamic asymmetry of the universe
in which we live. This explanation does not presuppose an objective {\it
causal}%
asymmetry, however.%
} Thus the asymmetry we think we see in the photon case is explained
as a shadow of a real asymmetry elsewhere, and no longer seems
in conflict with the symmetry of the photon's intrinsic physics.

\setlength{\parindent}{12pt}%
\setlength{\parskip}{0pt} In the present context, another advantage
of this suggestion might seem to be that it promises to explain
why the backward causation interpretation of Bell's results meets
with such resistance. If ordinary causal talk is a product of
some very basic characteristics of the sort of creatures we are,
small wonder that it is so deeply entrenched, and that we find
it so difficult to revise. This argument needs to be handled
with care, however, for if causal asymmetry is anthropocentric,
how can there be an objective issue as to whether causation is
``forward'' or ``backward''? The objective content of the backward
causation strategy seems to have been lost.

 We shall be able to turn this point to our advantage, however.
The view that the asymmetry of dependency is anthropocentric
will enable us to clarify the objective core of the advanced
action proposal. The trick will be to distinguish between two
kinds of correlational structures for the world. From the standard
point of view one possible structure ``looks like'' it simply
contains ordinary forward causation, while the other ``looks
like'' it contains a mixture of forward and backward causation.
The latter structure is the one we need to solve Bell's riddle.
Its objective content lies in its correlational structure, but
the causal viewpoint provides a useful way to distinguish it
from the classical alternative. All the same, it seems that Nature
has played a practical joke at this point, hiding the narrow
path to the advanced action interpretation between two much more
clearly marked alternatives: on the one hand the path that accepts
that our intuitive notion of directed causation is of fundamental
physical significance; on the other the path that banishes causation
from physics altogether. There is a middle way, but we shall
need to tread carefully.\footnote{I
suspect that the failure to notice this middle way is one reason
why the backward causation approach has been unpopular among
physicists. One commentator who seems to miss it is Bernard d'Espagnat.
In discussions of the advanced action interpretation advocated
by Costa de Beauregard (see note 40), d'Espagnat appears to move
from an acknowledgment of the ``apparently irreplaceable role
of man-centred concepts in the very definition of the causality
concept'' to the view that there is little of any novelty or
promise in the claim that quantum mechanics reveals backward
causation. See his (1989a), pp.\ 229-31, and (1989b), pp.\ 144-5.
}

\setlength{\parindent}{12pt}%
\setlength{\parskip}{0pt} So the problem of the asymmetry of
dependency---the puzzle exemplified by the photon case---promises
an illuminating route to a deeper understanding both of what
is at issue in the debate about backward causation in quantum
mechanics, and of the place of our ordinary views about causation
in our system of theoretical commitments as a whole. And that
in turn promises to counter the conceptual inertia we noted at
the beginning of the section. In so far as that inertia is rationally
grounded, much of it seems to stem from uncertainty concerning
the physical content of the advanced action interpretation; an
uncertainty engendered by the philosophical complexities of the
relationship between causation and physics. By clarifying the
commitments of the interpretation in this way, we may thus hope
to diminish the grounds for rational conservatism.

 My treatment of these issues goes like this. The first task
is to outline a case for a treatment of the asymmetry of dependency
on the anthropocentric lines just indicated. The argument will
be couched in terms of counterfactual conditionals, which seem
to provide the most transparent route to the insights we are
after. As we shall see, a natural objection to this approach
is that in tying the asymmetry of dependency to a conventional
feature of our use of counterfactuals, it fails to capture the
intuitive objectivity of the asymmetry. Responding to this objection
will lead us back to the issue as to whether the approach leaves
room for backward causation, and hence to the main concerns of
the paper. I shall argue that the approach leaves just the space
that the advanced action interpretation requires, and show how
it clarifies what is objectively at issue between this interpretation
and its more conventional rivals. Finally, we'll see that this
clarification puts us in range of a powerful new argument for
the interpretation, namely that it seems to respect a temporal
symmetry in microphysics that the classical view is bound to
deny.

\begin{center}%
\setlength{\parskip}{12pt}%
{\it 8. Asymmetry conventionalised.}

\end{center}%
\setlength{\parskip}{0pt}The intuition I appealed to above was
that the state of our photon does not depend on the fact that
it is going to pass through a polarising lens in the near future,
but may depend on the fact that it passed through such a lens
in the distant past. What makes this intuition puzzling is that
it is temporally asymmetric, without there being any apparent
basis for such an asymmetry in the physical phenomena themselves.

 A useful approach to this puzzle is to do away with explicit
talk of dependency, by couching the intuitive asymmetry in counterfactual
terms. Intuitively, it seems true that
\begin{itemize}
\item[(1)]
 If the distant polariser had been differently oriented the incoming
photon might now be in a different state (or might not have arrived
here at all);
\end{itemize}
and yet false that
\begin{itemize}
\item[(2)]
If the local polariser had been differently oriented, the state
of the incoming photon might have been different.
\end{itemize}%
Granted that this contrast seems to capture the intuitive asymmetry,
a tempting suggestion is that the asymmetry doesn't really have
anything to do with the physical entities themselves, but is
buried in the content of counterfactual claims. After all, a
popular account of the semantics of counterfactuals goes something
like this. When we assess a counterfactual of the form ``If X
had happened at $t$%
{\normalsize $_{1}$} then Y would have happened at $t$%
{\normalsize $_{2}$}'' we consider the consequences of a hypothetical
alteration in the course of events. We ``hold fixed'' the course
of events %
{\it prior} to $t$%
{\normalsize $_{1}$}, assume that X occurs at that time, and
consider the course of events that %
{\it follows}. Roughly speaking, we take the conditional to
be true if Y at $t$%
{\normalsize $_{2 }$} is a consequence of the joint assumption
of the actual history prior to $t$%
{\normalsize $_{1}$} and the occurrence of X at $t$%
{\normalsize $_{1}$}.\footnote{See
Jackson (1977), for example.
}%

\setlength{\parindent}{12pt}%
\setlength{\parskip}{0pt} According to this account, (1) says
that if we hold fixed the history of the photon before it reaches
the distant polariser, but assume that polariser to be differently
oriented, we may derive different consequences for the state
of the photon before it reaches the Earth (i.e., in the region
between the two polarisers). Whereas (2) says that if we hold
fixed the history of the photon before it reaches the %
{\it local} polariser, and assume that polariser to be oriented
differently, we may derive different consequences for the state
of the photon before it reaches the local polariser. The latter
claim is clearly false, but for a rather trivial reason: the
state of the photon doesn't change under the assumed circumstances,
because the assumption includes the supposition that it doesn't!\footnote{(2)
isn't really the temporal image of (1), in other words, and so
the difference in their truth values doesn't constitute a genuine
temporal asymmetry.
}

\setlength{\parindent}{12pt}%
\setlength{\parskip}{0pt} The standard semantics for counterfactuals
thus dissolves the puzzle of the photon case. Why doesn't the
present affect the past? Because to say that the present affected
the past would just be to say that if the present had been different
the past would have been different. And that in turn is just
to say that if we suppose the present to be different, while
the past remains the same, it will follow that the past is different.
This is untrue, of course, but simply on logical grounds. No
physical asymmetry is required to explain it.

 But is this the right way of dissolving the tension? An obvious
objection is that it makes our ability to affect the future but
not the past a conventional or terminological matter. After all,
couldn't we have adopted the opposite convention, assessing counterfactuals
by holding the future fixed---in which case wouldn't we now be
saying that present events depend on the future but not the past?
And if there is this alternative way of doing things, how could
one way be %
{\it right} and other way %
{\it wrong}? How could it be objectively the case, as it seems
to be, that we can affect the future but not the past?

 This objection might have a second aspect to it, closely related
to our present concerns: we might feel that even if the past
doesn't depend on the future, it isn't %
{\it logically} impossible that it should do so, in the way
that this diagnosis would suggest. The suggested account of counterfactuals
thus seems in one sense too weak, in failing to give due credit
to a genuine difference between the past and the future; and
in another sense too strong, in ruling out backward dependency
by fiat.

 In the next section I shall sketch what seems to me to be an
adequate response to this objection, and try to show how it illuminates
our main concern.\footnote{A
thorough response to the objection would involve a critique of
alternative accounts of the asymmetry of dependency. I shall
not attempt this here, but see my (1992) and (1993).
} The latter aspect of the objection is the more directly relevant
of the two, of course, but the former is the more crucial to
the credentials of the anthropocentric approach to asymmetry
as a whole, and needs to be tackled first.

\begin{center}%
\setlength{\parindent}{12pt}%
\setlength{\parskip}{12pt}%
{\it 9. Convention objectified.}

\end{center}%
\setlength{\parskip}{0pt}The objection is that the conventionalist
account would make the asymmetry of dependency less objective
than it actually seems to be---a matter of choice, in effect.
The appropriate response, I think, is to draw attention to the
constraints on this ``choice'' imposed by the circumstances in
which we find ourselves, and over which we have no control. Because
these constraints are so familiar we fail to notice them, and
mistake the asymmetry they impose for an objective feature of
the world. The moral is that things may %
{\it seem} more objective than they actually are, when the source
of the subjectivity is not an obvious one.

 The main constraint turns on the fact that counterfactuals are
used in %
{\it deliberation}, and that deliberation is a temporally asymmetric
process. Considered as structures in spacetime, agents (or ``deliberators'')
are asymmetric, and thus orientable along their temporal axis.
The crucial point is that this orientation is not something an
agent is able to choose or change. From an agent's point of view
it is simply one of the ``givens'', one of the external constraints
within which she or he operates.

 Thus although our temporal orientation is probably a contingent
matter---there might be agents with the opposite temporal orientation
elsewhere in the universe---it is not something that we can change.
Given what we do with counterfactuals in deliberation (more on
this in a moment), we thus have no option but to assess them
the way we do---that is, to hold the past rather than the future
fixed. It is objectively true that %
{\it from our perspective}, we can't affect our past. Unable
to adopt any other perspective, however, and failing to notice
the relevance and contingency of our temporal orientation, we
fail to see that what we have here is a relational truth, a truth
about how things are from a perspective, rather than absolute
truth about the world. The account thus provides a kind of quasi-objectivity.
It explains why we think of the asymmetry of dependency as an
objective matter.

 The account relies on the claim that agents are asymmetric and
thus orientable in time. This doesn't seem particularly contentious,
but it would be useful to have a better understanding as to what
this asymmetry involves. I think there are two ways to approach
the issue. One would seek to characterise the asymmetry in terms
of a formal model of deliberation. Roughly speaking, the goal
would be to map the structure of an ideal deliberative process---to
map it from an atemporal perspective, laying out the steps %
{\it en bloc}---and hence to be able to point to an intrinsic
asymmetry along the temporal axis. This seems a plausible project,
involving little more than transcribing standard dynamic models
of deliberation into an atemporal key.

 The other approach would be a phenomenological one, going something
like this. From the inside, as it were, we perceive a difference
between the inputs and the outputs of deliberation---between
``incoming information'', which appears as ``fixed'', or ``given'',
and ``outgoing behaviour'', which appears as ``open'', or subject
to our control. The contrast is a subjective one, a feature of
what it %
{\it feels like} to be an agent, so that such difference would
not be apparent from a genuinely external perspective. An atemporal
God would just see a pattern of correlations in both temporal
directions between the deliberator's internal states and various
environmental conditions (and wouldn't regard the temporal ordering
implied by the terms ``input'' and ``output'' as having any objective
significance, of course).\footnote{This
phenomenological account will appeal more than the formal one
to philosophers who seek to ground folk concepts in folk experience.
Do we need to adjudicate between the two? I don't think so. It
seems reasonable to expect that they will turn out to be complimentary.
In effect, the formal approach will simply be describing the
internal structure of the pattern of correlations on which the
phenomenological approach depends.
}

\setlength{\parindent}{12pt}%
\setlength{\parskip}{0pt} Either way, where do counterfactuals
fit into the picture? We need to think about the role of counterfactuals
in deliberation. An agent normally has the choice of a number
of options, and bases her choice not on the desirability of the
immediate options themselves, but on that of what might be expected
to follow from them. Thus a typical deliberative move is to take
what is ``given'' or ``fixed''---or rather, in practice, what
is known of what is taken to be fixed---and then to add hypothetically
one of the available options, and consider what follows, in accordance
with known general principles or laws. The temporal orientation
of this pattern of reasoning follows that of the agent's perspective.
Broadly speaking, what we hold fixed when assessing counterfactuals
according to the standard semantics is what presents itself to
us as fixed from our perspective as agents. So long as counterfactuals
are to maintain their association with deliberation,\footnote{There
is no necessity in this, of course. We could, and in practice
certainly do, use counterfactuals for other purposes. The claim
is simply that in those contexts in which the symmetry of dependency
seems vivid to us---contexts such as that of (1) and (2) above---the
use is one which depends on this connection with deliberation.
} in other words, our choice of the ``hold the past fixed'' convention
is governed by the contingent but nevertheless unchangeable fact
of our orientation as agents. We have no choice in the matter.

\setlength{\parindent}{12pt}%
\setlength{\parskip}{0pt} This account explains the %
{\it apparent} objectivity of the asymmetry of dependency, and
thus meets the charge that the conventionalist strategy makes
the asymmetry too weak. However, the objection was that the strategy
is also too strong, in making it analytic and hence %
{\it a priori} that we cannot affect the past. We now need to
show that the conventionalist approach leaves a loophole for
backward causation---in other words, that this anthropocentric
account to the asymmetry of dependency does not backfire on the
main project of the paper.\footnote{Note
that what we are interested in showing is that it is an empirical
possibility that the world might contain what
{\it we}  would describe as backward causation; not merely
that there might be differently-oriented creatures who would
see it as containing what we would have to describe---to the
extent that we could describe it---as backward causation. The
issue is whether the conventionalist proposal make sense of the
idea that even
{\it from our own perspective} it is
{\it a posteriori} that we can't affect the past.
}

\vspace{24pt}

\begin{center}%
\setlength{\parindent}{12pt}%
\setlength{\parskip}{12pt}%
{\it 10. Backward dependency legitimised.}\nopagebreak[4]
\end{center}\nopagebreak[4]%

\nopagebreak[4]
\setlength{\parskip}{0pt}
\setlength{\parindent}{0pt}%
The crucial step is to appreciate that
although it is %
{\it a priori} is that we can't affect what we know about at
the time of deliberation, it is %
{\it a posteriori} that what we can get as input to deliberation
is all and only what lies in our past. It is an interesting question
why we should assume that this is the case. Even if experience
teaches us that what we know about via memory and the senses
always lies in the past, we would be affirming the consequent
to conclude that anything that lies in the past is something
that might in principle be known about, and hence something inaccessible
to deliberation. In fact, it seems that the relationship between
temporal location and epistemological accessibility is not only
contingent (in both directions), but rather under determined
by our actual experience. For all that our limited experience
tells us, there might actually be some of the past to which we
do not have access, and perhaps some of the future to which we
do. The epistemological boundaries seem to be an empirical matter,
therefore, to an extent that ordinary practice tends to overlook.

\setlength{\parindent}{12pt}%
Here the point connects with our earlier discussion. We noted
in section 4 that as Dummett (1964) points out, that it is possible
to avoid the bilking argument so long as one confines oneself
to the claim that one can affect bits of the past which are epistemologically
inaccessible at the time that one acts. To make the example concrete,
suppose it were suggested that the present state $\phi$ of our incoming
photon depended on the setting of the local polariser in its
future, as well as on that of the distant polariser in its past.
In this case the bilking argument would require that we measure
$\phi$ before the photon arrives, and thereby set the future polariser
to conflict with the claimed correlation. So if the only way
to measure $\phi$ is to pass the photon through a polariser, and the
claim in question is simply that the value of $\phi$ depends on the
orientation of the %
{\it next} polariser it encounters, bilking becomes impossible.
Interposing another measurement would mean that the %
{\it next} measurement is not the one it would have been otherwise,
so that the conditions of the claimed dependency would no longer
obtain. According to the claim under test, such a measurement
cannot be expected to reveal the value that $\phi$ %
{\it would have had} if the measurement not been made.

 By slipping into talk of counterfactuals and backward dependency
here, I have tried to illustrate that this admission of limited
retro-dependency is not as alien as might be imagined. It seems
in particular that our use of counterfactuals is already somewhat
more flexible than the model we have been using suggests---already
sufficiently flexible to handle the kind of cases that Dummett's
loophole admits, in fact. What we seem to do in such cases is
just what Dummett would have us do, in effect: we assess counterfactuals
not by holding fixed everything prior to the time of the antecedent
condition, but by holding fixed what we have access to at that
time (without disturbing the background conditions of the claimed
correlation).

 This shows how the conventionalist can make sense of the possibility
of backward dependency (and thus meet the second part of the
objectivist's challenge). Note that this response does not require
that the ordinary use of counterfactuals is already unambiguously
committed to this possibility, in the sense that it is already
configured to take advantage of the Dummettian loophole. If that
were so the assumption that the past does not depend on the future
would surely be less deeply ingrained than it is in scientific
theorising. It seems to me more accurate to say that there is
a systematic indeterminacy in our ordinary notions of causal
and counterfactual dependency. Ordinary usage does not clearly
distinguish two possible conventions for assessing counterfactuals:
a mode under which the entire past is held fixed, and a mode
under which only the %
{\it accessible} past is held fixed. If we follow the former
convention then it is analytic that the state of the incoming
photon does not ``depend'' on the local polariser setting; whereas
if we follow the latter convention it may do so. But this difference
reflects an issue about how we should use the terms involved,
rather than a disagreement about the objective facts of the matter.

 The nice thing about this point is that it suggests a natural
way of characterising the objective core of the advanced action
interpretation. For it is clear that advocates of the two conventions
for counterfactuals might agree on the relevant categorical facts.
In the photon case, for example, they might agree that there
is a one-to-one correlation between some feature of the state
of the incoming photon and the setting of the later
polariser.\footnote{Strictly
speaking what they will agree on is that this correlation holds
in the class of actual cases of this kind. Modal generalisations
might prove contentious.
} The difference is simply that the second convention (``hold
the whole past fixed'') precludes us from interpreting this correlations
in terms of backward influence.

\setlength{\parindent}{12pt}%
\setlength{\parskip}{0pt} Thus by separating the issue of the
correlational structure of the world from that of the appropriate
convention for counterfactuals, we seem to be able to insulate
the empirical question we are interested in---a question concerning
the structure of the microworld---from the heat of folk intuitions
about what might depend on what. Faced with an opponent who claims
to find backward dependency incoherent, we can simply concede
counterfactual practice, and fall back to the categorical issue.

 The point of this tactical move is to enable us to keep our
eye on the issue that really matters for the interpretation of
quantum mechanics, and it doesn't commit us to being even-handed
on the issue as to which choice of convention for counterfactuals
would be the more appropriate in a world with correlations of
the imagined kind. It seems to me that there would good reasons
for preferring the weaker convention (``hold fixed the accessible
past''), which become evident if we consider what the stronger
convention would have us say about the imagined case. To simplify
as much as possible, suppose that there is agreed to be a strict
correlation between a particular polariser setting S and a particular
earlier state $\phi$%
{\normalsize $_{s}$} of the incoming photon. If counterfactuals
are assessed according to the stronger convention, it cannot
be the case that both the following counterfactuals are true
(for if we hold fixed events before S, then one consequent or
other is bound to be false, even under the counterfactual supposition
in question).

\begin{itemize}

\item[(3)] If we were to bring about S, this would ensure that $\phi$%
{\normalsize $_{s}$}

\item[(4)]
If we were to bring about not-S, this would ensure that not-$\phi$%
{\normalsize $_{s}$}.

\end{itemize}%
\setlength{\parindent}{12pt}

 Suppose for the sake of argument that the incoming photon is
not in state $\phi$%
{\normalsize $_{s}$}, so that it is (3) that has a false consequent
(under the assumption that the past is held fixed). The two possibilities
seem to be to regard (3) as false or to regard it as somehow
meaningless, or otherwise inappropriate. To go the first way
is to say that the agreed correlation between S and $\phi$%
{\normalsize $_{s}$} does not support counterfactuals. To go
the second way seems to be to say that means--end reasoning breaks
down here---that it doesn't make sense to suppose we might do
S. Neither course seems particularly satisfactory. We are supposing
that all parties acknowledge that as a matter of fact the correlation
does always obtain between S and $\phi$%
{\normalsize $_{s}$}, including on whatever future occasions
there might happen to be on which we bring about S. Outcomes
of actual actions thus assured, it is hard to see how the refusal
to acknowledge the corresponding counterfactuals could seem anything
but wilful---so long, at any rate, as we claim the ability to
bring about S at will. Denying free will seems to be an alternative,
but in this case it should be noted that the phenomenology isn't
going to be any different. So there is nothing to stop us from
going through the motions of deliberating, as it were. Within
this scheme of quasi-deliberation we'll encounter quasi-counterfactuals,
and the question as to how these should be assessed will arise
again. Hold fixed the past, and the same difficulties arise all
over again. Hold fixed merely what is accessible, on the other
hand, and it will be difficult to see why this course was not
chosen from the beginning.

 Thus I think that if we were to find ourselves in a world in
which the notions of the past and the epistemologically accessible
came apart in this way---a world in which it thus became important
to resolve the ambiguity of current counterfactual usage---a
resolution in favour holding fixed merely what is accessible
would be the more satisfactory. For present purposes, however,
the important point is that this issue about how we should use
counterfactuals is quite independent of the empirical issue as
to whether the world has the correlational structure that would
require us to make the choice.

 In sum, the proposal that the asymmetry of dependency rests
on a conventional asymmetry in the content of counterfactual
claims leaves open the possibility of exceptional cases, in which
the past would indeed be properly said to depend on the future.
But it leaves open this possibility not in the direct sense that
current usage unambiguously admits it, but in the sense that
in conceivable physical circumstances the most natural way to
clarify and disambiguate current usage would be such as to recognise
such backward influence. Finally, it should be emphasised that
this sort of backward causation will not be the time-reverse
of ordinary %
{\it forward} causation, since our own temporal orientation
and perspective remains fixed.

 Arguing for a bare possibility is one thing. Arguing that the
actual world is like this is quite another, of course, but it
is worth noting that any such argument seems bound to be indirect.
The kind of backward influence concerned is not likely to be
directly observable case by case, for the simple reason that
if it were, the bilking argument would again gain a foothold.
So the case for such a model of the world will have to rely on
non-observational considerations---considerations of simplicity,
elegance and symmetry, for example. This takes us neatly back
to our starting point. The argument for the advanced action interpretation
rested on a number of theoretical advantages of this kind---especially
its ability to avoid non-locality and hence conflict with special
relativity, but also the promise of the advantages of ``incompleteness''
interpretations in general. It is useful to be reminded that
we shouldn't expect evidence of a more direct kind: we shouldn't
expect to ``see'' backward influence in action, as it were.

\begin{center}%
\setlength{\parskip}{12pt}%
{\it 11. The symmetry argument for advanced action.}%

\end{center}%
\setlength{\parskip}{0pt}But is the indirect case as strong
as it might be? A tempting thought is that there might be a symmetry
argument in favour of the advanced action view. If classical
views of microphysics turned out to embody a temporal asymmetry
which could be removed by reinterpreting the theories concerned
in such a way as to allow backward dependency, then we would
seem to have a strong argument in favour of
such a reinterpretation.\footnote{Why?
Simply because it would eliminate an otherwise mysterious breach
of symmetry. Reasoning of this kind is common in physics.
}

\setlength{\parindent}{12pt}%
\setlength{\parskip}{0pt} It is important to note that the simple
asymmetry of dependency exemplified by the photon case is not
problematic in this way, being explicable in terms of the conventional
asymmetry of counterfactuals. Recall that the crucial point was
that the counterfactuals (1) and (2) are not genuinely the temporal
inverse of one another, so that their difference in truth value
does not require a genuine temporal asymmetry. What would be
problematic would be the claim that the categorical constituents
of the world were asymmetric on the micro scale. It has sometimes
been noted that this seems to be true of the standard interpretation
of quantum mechanics, according to which the wave function is
localised %
{\it after} but not %
{\it before} a measurement interaction.\footnote{See
Penrose (1989), pp.\ 354-6, for example.
} So much the worse for the standard interpretation, in my view,
but that's an argument for another time. In the present context
the interesting thing is that any hidden variable theory which
fails to allow advanced action seems likely to violate this symmetry
requirement.

\setlength{\parindent}{12pt}%
\setlength{\parskip}{0pt} To see why this is so, let us return
to the photon case. The easiest way to think about the difference
between the advanced action interpretation and its orthodox rivals
is in terms of counterfactuals assessed according to the ``hold
fixed the accessible past'' convention, which permit retro-dependency.
(There is no logical impropriety in discussing the case in these
terms, so long as we keep in mind that the objective content
of the issue concerns patterns of correlation.) The orthodox
view is that the state $\phi$ of the photon between the polarisers
does not depend on the setting of the future polariser. If we
assume that what is accessible is the state of the photon in
the region prior to the past polariser, this independence assumption
amounts to the following:

\vspace{3pt}
\begin{itemize}
\item[(5)]
With the history prior to the past polariser held fixed, changes
in the setting of the future polariser do not imply changes in
the value of $\phi$ in the region between the polarisers.
\end{itemize}%
\vspace{3pt}
In order to determine
whether this view involves a temporal asymmetry, we need to ask
whether it also endorses the temporal inverse of this assumption,
which is:

\vspace{3pt}
\begin{itemize}
\item[(6)]
With the course of events after the photon passes the %
{\it future} polariser held fixed, changes in the setting of
the %
{\it past} polariser do not imply changes in the value of $\phi$
in the region between the polarisers.
\end{itemize}%

\vspace{3pt}
\setlength{\parindent}{12pt}There is little intuitive
appeal in this latter proposition, however. The easiest way to
see this is to imagine more familiar cases in which we talk about
alternative histories. For example, suppose that we have an artefact,
known to be a year old, which could have been manufactured by
one of two processes. Given its present condition, does our view
about its condition say six months ago depend on our view about
its origins? Obviously it might well do so. In the one case,
for example, the distinctive patina on the surface of the object
might have been a product of the manufacturing process itself;
in the other case it might have been acquired gradually as the
object aged. So there is no reason to expect the condition of
the object in the intervening period to be independent of what
happened to it in the past.

 In endorsing (5) but not (6), then, the orthodox view seems
to be committed to a genuine asymmetry---the sort of asymmetry
which is %
{\it not} needed to account for the contrast between (1) and
(2). The advanced action view avoids this asymmetry, by rejecting
(5) as well as (6). This appears to be a powerful additional
argument in favour of the advanced action view. How could such
an argument have been overlooked? The answer, I take it, is that
it has not been noticed that the orthodox assumption involves
anything more than the ordinary asymmetry of physical dependency,
exemplified by the contrast between (1) and (2). Since the latter
contrast is rightly assumed to involve no violation of physical
symmetry principles, it has been taken for granted that the same
is true in the cases of interest to quantum mechanics.

\vspace{12pt}

\begin{center}%
\setlength{\parskip}{12pt}%
\nopagebreak {\it 12. Summary: the objective core of the advanced action
interpretation.}\nopagebreak%

\nopagebreak
\end{center}%
\setlength{\parskip}{0pt}Understanding the anthropocentric origins
of the asymmetry of dependency is thus an important step towards
appreciating the attractions and physical significance of an
advanced action interpretation of quantum mechanics. On the one
hand it enables us to see that despite the conventional character
of this asymmetry, there is a real physical issue at stake. (Hence
it counters the tendency
\footnote{Cf.
note 25
} to dismiss the approach as being either ``metaphysical'' or
``subjective'', and hence of no relevance to physics.) On the
other hand, by unravelling some the connections between our folk
intuitions about causation and our place in the physical world,
it helps us to see that these intuitions do not provide an authoritative
guide to the issue in quantum mechanics.

\setlength{\parindent}{12pt}%
\setlength{\parskip}{0pt} In particular, the recognition of
the subjectivity of causal asymmetry does not throw the baby
out with the bath water, as we put it earlier, leaving no objective
content to the advanced action view. The problem was simply that
we didn't have a clear impression of the distinction between
the water and the baby---that is, between our subjectively-grounded
intuitions concerning causal asymmetry and an issue concerning
patterns of correlation in the world. Once the distinction has
been drawn, however, it is possible with a little care to discard
the dirty water and keep the baby.

 The baby is a proposition concerning the correlational structure
of the microworld, and we have seen that there are two very different
forms this structure might take. There is an objective distinction
between worlds which look as if they contain unidirectional causation
and worlds which look as if they contain bidirectional causation
(where ``look as if'' is to be filled out in terms of the conventionalist's
account of how the temporally asymmetric nature of agents comes
to be reflected in their concept of causation). With the benefit
of hindsight we can see that we have never had good reason to
exclude the structure that permits interpretation in bidirectional
terms. In order to be able to see this, however, it was essential
to appreciate the subjective character of the interpretation.\footnote{Compare:
with hindsight we see that we really had no good reason to expect
microphysical objects to be coloured, but in order to appreciate
that this is the case, we first had to appreciate the subjective
nature of colour concepts.
}%
{\normalsize  }

\begin{center}%
\setlength{\parindent}{12pt}%
\setlength{\parskip}{12pt}%
{\it 13. TRIPLET SHOCK LATEST!}%

\end{center}%
\setlength{\parskip}{0pt}As some readers will already be aware,
there is an interesting new class of Bell-like results in quantum
mechanics, which seem to achieve Bell's conclusions by non-statistical
means. It is natural to wonder whether these new results---the
Greenberger-Horne-Zeilinger (GHZ) cases, as they have become
known---have a bearing on the arguments of this paper. To my
knowledge, the GHZ argument has not yet been expounded in non-technical
terms in the general philosophical literature.\footnote{The
most accessible expositions are those of Mermin (1990) and Maudlin
(1994), pp.\ 24-8. My account draws on that of Clifton, Pagonis
and Pitowsky (1992).
} By a stroke of good fortune, however, it turns out that here
too Doppelg\"{a}nger has been here first. In recently de-classified
research,\footnote{Allegedly
funded by the Copenhagen Interpretation Authority.
} Doppelg\"{a}nger conducted further analyses of the Ypiarian
police records, obtaining results strikingly parallel to those
predicted by GHZ. The purpose of this section is to provide a
brief exposition of these results, so as to show that the GHZ
results simply add weight to the existing case for advanced action.

\setlength{\parindent}{12pt}%
\setlength{\parskip}{0pt} Doppelg\"{a}nger's clandestine investigations
led him from twins to triplets. To his surprise, he found that
the Ypiarian criminal code embodied special exemptions for triplets,
who were excused for adultery on grounds of diminished childhood
responsibility. When subject to random interrogation, then, triplets
were only asked one of the two questions Are you a thief? and
Are you a murderer? In records of these interrogations, Doppelg\"{a}nger
found that on occasions on which all three members of a set of
triplets were questioned in this way, their answers always conformed
to the following pattern:

\begin{description}

\item[{\sc same*}]%
When all three triplets were asked the first question,
an odd number of them said ``no''.%
\item[{\sc diff*}]%
When two triplets were asked the second question and one
the first, an even number (i.e. two or none) of them said ``no''.
\end{description}%
\setlength{\parindent}{12pt}As in the twins case,
Doppelg\"{a}nger asked himself whether these results could be
explained in terms of ``local hidden variables''---that is, in
terms of psychological factors predisposing each triplet to respond
in a certain way to either of the possible questions, independently
of the concurrent experiences of his or her fellow triplets.
Reasoning as follows, he decided that this was impossible.

 Suppose that there are such psychological factors. Given a particular
set of triplets, let us write %
{\it x}%
{\it %
{\normalsize $_{1}$}} and %
{\it y}%
{\it %
{\normalsize $_{1}$}} for the factors responsible for the answer
the first triplet would give to the first and second question,
respectively, and similarly %
{\it x}%
{\it %
{\normalsize $_{2}$}}, %
{\it y}%
{\it %
{\normalsize $_{2}$}}, %
{\it x}%
{\it %
{\normalsize $_{3}$}} and %
{\it y}%
{\it %
{\normalsize $_{3}$}} for the corresponding factors in the second
and third triplets. And let us think of these factors as having
values $+1$ or $-1$, according to whether they dispose to a positive
or negative answer. Then %
{\sc same*} implies that the product of %
{\it x}%
{\it %
{\normalsize $_{1}$}}, %
{\it x}%
{\it %
{\normalsize $_{2}$}} and %
{\it x}%
{\it %
{\normalsize $_{3}$}} is $-1$ (since it contains an odd number
of negative factors); and %
{\sc diff*} implies that each of the products %
{\it x}%
{\it %
{\normalsize $_{1}$}}%
{\it y}%
{\it %
{\normalsize $_{2}$}}%
{\it y}%
{\it %
{\normalsize $_{3}$}}, %
{\it y}%
{\it %
{\normalsize $_{1}$}}%
{\it x}%
{\it %
{\normalsize $_{2}$}}%
{\it y}%
{\it %
{\normalsize $_{3}$}}%
{\normalsize $_{ }$} and %
{\it y}%
{\it %
{\normalsize $_{1}$}}%
{\it y}%
{\it %
{\normalsize $_{2}$}}%
{\it x}%
{\it %
{\normalsize $_{3}$}}%
{\normalsize $_{ }$} has value $+1$ (since it contains an even number
of negative factors). Taken together these results in turn imply
that the combined product

\begin{center}
 $($%
{\it x}%
{\it %
{\normalsize $_{1}$}}%
{\it x}%
{\it %
{\normalsize $_{2}$}}%
{\it x}%
{\it %
{\normalsize $_{3}$}})(%
{\it x}%
{\it %
{\normalsize $_{1}$}}%
{\it y}%
{\it %
{\normalsize $_{2}$}}%
{\it y}%
{\it %
{\normalsize $_{3}$}})(%
{\it y}%
{\it %
{\normalsize $_{1}$}}%
{\it x}%
{\it %
{\normalsize $_{2}$}}%
{\it y}%
{\it %
{\normalsize $_{3}$}})(%
{\it y}%
{\it %
{\normalsize $_{1}$}}%
{\it y}%
{\it %
{\normalsize $_{2}$}}%
{\it x}%
{\it %
{\normalsize $_{3}$}}$) = (-1)(+1)(+1)(+1) = -1.$
\end{center}
This is impossible, however, since each individual factor occurs
exactly twice on the left hand side, so that negative factors
must cancel out. Hence local hidden variables cannot account
for the observed results.

 Our present interest in this case lies in the fact that the
triplet correlations that Doppelg\"{a}nger discovered in the
Ypiarian case exactly match those predicted by the recent GHZ
results in quantum mechanics. The behaviour of Ypiarian triplets
parallels that of sets of three spin-$1/2$ particles in the so-called
GHZ state, when subject to combinations of spin measurements
in one of two chosen directions orthogonal to their lines of
flight. Particle 1 can thus have its spin measured in direction
{\it x}%
{\it %
{\normalsize $_{1}$}} or direction %
{\it y}%
{\it %
{\normalsize $_{1}$}}, particle 2 in direction %
{\it x}%
{\it %
{\normalsize $_{2}$}} or %
{\it y}%
{\it %
{\normalsize $_{2}$}}  and particle 3 in direction %
{\it x}%
{\it %
{\normalsize $_{3}$}} or %
{\it y}%
{\it %
{\normalsize $_{3}$}}. An argument exactly parallel to the one
just given shows that a local hidden variable theory cannot reproduce
the predictions of quantum mechanics concerning combinations
of such measurements.

 Notice that unlike Bell's Theorem, the argument is combinatorial
rather than statistical in character. Like Bell's Theorem, however,
the GHZ argument depends on the Independence Assumption. That
is, it requires the assumption that the values of hidden variables
do not depend on what is to happen to the particles in question
in the future---in particular, on the settings of future spin
measurements. In the presentation of the argument by Clifton,
Pagonis and Pitowsky (1992) for example (whose terminology I
have borrowed above, in describing the Ypiarian parallel), the
assumption is introduced in the following passage:

\begin{quotes}
{[W]}e need to assume that $\lambda$ [the set of hidden variables] is
compatible
with all four measurement combinations. This will be so if: (a)
the settings are all fixed just before the measurement events
occur, so that  $\lambda$ lies in the backwards light-cones of the setting
events, and (b) the choice of which settings to fix is determined
by \mbox{(pseudo-)random} number generators whose causal history is
sufficiently disentangled from  $\lambda$. (1992, p.~117)
\end{quotes}%
The advanced action interpretation
will simply reject proposition (a), of cour\-se.\footnote{Proposition
(b) serves to exclude the alternative path to rejection of the
Independence Assumption which we mentioned in note 4.%
} In the present context the important point is that the GHZ
argument provides no new argument against the advanced action
interpretation. On the contrary, if anything, it simply adds
impressive new weight to the view that the alternatives to advanced
action are metaphysically unpalatable.

\begin{center}%
\setlength{\parindent}{12pt}%
\setlength{\parskip}{12pt}%
{\it 14. Conclusion.}
\nopagebreak
\end{center}%
\setlength{\parskip}{0pt}The metaphysical temptation that Bell
noticed is a much better offer than he himself took it to be.
It doesn't raise any new difficulties for free will, or any awkward
causal paradoxes. Its objective core seems to boil down to a
rather harmless proposal concerning the correlational structure
of the microworld---indeed, a proposal which seems independently
attractive on symmetry grounds. Yet it has the advantages that
Bell seems to have recognised: it offers us a route to an Einsteinian
realism about quantum mechanics which is local, and therefore
compatible with special relativity; and which shares the well
known advantages of the Einsteinian view that quantum mechanics
is incomplete. It is time that this neglected approach received
the attention it so richly deserves.\footnote{A
brief guide to further reading: the earliest, best known and
certainly the most prolific advocate of an advanced action interpretation
is O.\ Costa de Beauregard, whose papers on the topic date back
to 1953. His more recent papers include (1977) and (1979). (Costa
de Beauregard's views are discussed by d'Espagnat in the works
mentioned in note 25.) The most highly developed version of the
interpretation seems to be that of J.\ G.\ Cramer (1986, 1988).
In the former paper (pp.\ 684-5) Cramer compares his view to earlier
advanced action interpretations. The theoretical basis of Cramer's
approach seems to me to be questionable, however: see my (1991b),
pp.\ 973-4. Other advocates of advanced action in quantum mechanics
include Davidon (1976), Rietdijk (1978), Schulman (1986) and Sutherland (1983).
See also Price (1984).
}

\vspace{24pt}

\setlength{\parindent}{0pt}

\setlength{\parskip}{0pt}School of Philosophy

University of Sydney

Australia 2006

\setlength{\parindent}{12pt}%

\begin{center}%
{\bf %
{\it

\newpage
}}%
{\bf %
{\it %
{\Large References}}}

{\bf %
{\it }}

\end{center}%
\setlength{\parindent}{0pt}%
\setlength{\parskip}{3pt}%

Bell,
J.\ S. 1964: ``On the Einstein-Podolsky-Rosen Paradox'', %
{\it Physics}, 1, pp.\ 195-200.

---------1987: %
{\it Speakable and Unspeakable in Quantum Mechanics}, Cambridge:
Cambridge University Press.

Bohm, D., 1951: %
{\it Quantum Theory}, Englewood Cliffs, N.J.: Prentice-Hall.

Butterfield, J., 1994: ``Stochastic Einstein Nonlocality and
Outcome Dependence'', in Prawitz, D. and Westerdahl, D., eds.,
{\it LMPS91, Selected Papers from the Uppsala Congress}, Dordrecht:
Kluwer.

Clifton, R., Pagonis, C. and Pitowsky, I. 1992: ``Relativity,
Quantum Mechanics and EPR'', in Hull, D., Forbes, M. and Okruhlik,
K. , eds., %
{\it PSA 1992, Volume 1}, pp.\ 114-28.

Costa de Beauregard, O. 1977: ``Time Symmetry and the Einstein
Paradox'', %
{\it Il Nuovo Cimento}, 42B, 1, pp.\ 41-64.

---------1979: ``Time Symmetry and the Einstein Paradox -- II'',
{\it Il Nuovo Cimento}, 51B, 2, pp.\ 267-279.

Cramer, J.\ G. 1986: ``The transactional interpretation of quantum
mechanics'', %
{\it Reviews of Modern Physics}, 58, 3, pp.\ 647-687.

---------1988: ``An overview of the transactional interpretation
of quantum mechanics'', %
{\it International Journal of Theoretical Physics}, 27, 2, pp.\
227-236.

Davidon, W.\ C. 1976: ``Quantum Physics of Single Systems'', %
{\it Il Nuovo Cimento}, 36B, 1, pp.\ 34-40.

Davies, P.\ C.\ W. and Brown, J.R., eds. 1986: %
{\it The Ghost in the Atom}, Cambridge: Cambridge University
Press.

d'Espagnat, B., 1989a: %
{\it Reality and the Physicist}, Cambridge: Cambridge University
Press.

---------1989b: ``Nonseparability and the Tentative Descriptions
of Reality'', in Schommers, W., ed., %
{\it Quantum Theory and Pictures of Reality}, Springer-Verlag,
pp.\ 89-168.

Dummett, Michael 1964: ``Bringing About the Past'', %
{\it Philosophical Review}, 73, pp.\ 338-359.

Einstein, A., Podolsky, B. and Rosen, N. 1935: ``Can Quantum-Mechanical
Description Of Physical Reality Be Considered Complete'', %
{\it Physical Review},%
 47, pp.\ 777-780.

Gribbin, J. 1990: ``The Man Who Proved Einstein Was Wrong'',
{\it New Scientist}, 24 November 1990, 33-35.

Horwich, Paul 1975: ``On Some Alleged Paradoxes of Time Travel'',
{\it Journal of Philosophy}, 72, pp.\ 432-444.

---------1987: %
{\it Asymmetries in Time}, Cambridge, Mass.: MIT Press.

Jackson, F. 1977: ``A Causal Theory of Counterfactuals'', %
{\it Australasian Journal of Philosophy}, 55, 3-21.

Lewis, David 1976: ``The Paradoxes of Time Travel'', %
{\it American Philosophical Quarterly}, 13, pp.\ 145-152.

Lockwood, Michael. 1989: %
{\it Mind, Brain and the Quantum: The Compound ``I''}, Oxford:
Basil Blackwell.

Maudlin, T. 1994: %
{\it Quantum Non-Locality and Relativity}, Oxford: Basil Blackwell.

Mermin, N.\ David 1981: ``Quantum Mysteries for Anyone'', %
{\it Journal of Philosophy}, 78, pp.\ 397-408.

---------1985: ``Is The Moon There When Nobody Looks? Reality
and Quantum Theory'', %
{\it Physics Today}, 38, 4, pp.\ 38-47.

---------1990: ``What's Wrong With These Elements of Reality?'',
{\it Physics Today}, 43, 6, pp.\ 9-11.

Penrose, Roger 1989: %
{\it The Emperor's New Mind}, Oxford: Oxford University Press.

Price, Huw 1984: ``The Philosophy and Physics of Affecting the
Past'', %
{\it Synthese}, 16, pp.\ 299-323.

---------1991a: ``Saving Free Will'', %
{\it New Scientist}, 12 January 1991, pp.\ 55-56.

---------1991b: ``The Asymmetry of Radiation: Reinterpreting
the Wheeler-Feynman Argument'', %
{\it Foundations of Physics}, 21, pp.\ 959-975.

---------1992: ``Agency and Causal Asymmetry'', %
{\it Mind} , 101, pp.\ 501-520.

---------1993: ``The Direction of Causation: Ramsey's Ultimate
Contingency'', in Hull, D., Forbes, M. and Okruhlik, K., eds.,
{\it PSA 1992, Volume 2}, pp.\ 253-267.

---------(forthcoming) ``Locality, Independence and the Pro-Liberty
Bell''.

Rietdijk, C.\ W. 1978: ``Proof of a Retroactive Influence'', %
{\it Foundations of Phys\-ics}, 8, 7/8, pp.\ 615-628.

Schulman, L.\ S. 1986: ``Deterministic Quantum Evolution through Modification
of the Hypothesis of Statistical Mechanics", {\it Journal of
Statistical Physics},
42, p.\ 689.

Sutherland, R.\ I. 1983: ``Bell's Theorem and Backwards-in-Time
Causality'', %
{\it International Journal of Theoretical Physics}, 22, 4, p.\
377-384.

Thom, Paul 1974: ``Time-Travel and Non-Fatal Suicide'', %
{\it Philosophical Studies}, 27, pp.\ 211-216.

\end{document}